\documentclass[oldversion]{aa}
\usepackage{amsmath}
\usepackage{txfonts}
\usepackage{natbib}
\usepackage{footnote}

\usepackage{graphicx}

\def\dex{\,{\rm dex}}

\def\url#1{{\tt#1}}

\begin{document}

\title{Evidence for a metal-poor population  in the inner Galactic  Bulge}

\author{M. Schultheis \inst{1}
  \and K. Cunha \inst{2}
\and G. Zasowski \inst{3,4}
\and A.~E.~Garc{\'i}a P\'erez \inst{5}
\and K.~Sellgren \inst{6}
\and V.~Smith \inst{7,2}
\and D.A.~Garc{\'i}a-Hern{\'a}ndez \inst{5,8}
\and O.~Zamora \inst{5,8}
\and T.~K.~Fritz\inst{9}
\and F.~Anders\inst{10,11}
\and C.~Allende Prieto\inst{5,8}
\and D.~Bizyaev\inst{12,13}
\and K.~Kinemuchi\inst{12} 
\and K.~Pan\inst{12}
\and E. ~Malanushenko\inst{12}
\and V. Malanushenko\inst{12}
\and M.D.~Shetrone\inst{14}
}

   \institute{ Universit\'e de Nice Sophia-Antipolis, CNRS, Observatoire de C\^ote d'Azur, Laboratoire Lagrange, 06304 Nice Cedex 4, France 
 e-mail: mathias.schultheis@oca.eu
\and 
 Observat'orio Nacional, Sao Crist\'ovao, Rio de Janeiro, Brazil
\and
Department of Physics \& Astronomy, Johns Hopkins University, Baltimore, MD 21218, USA
\and
NSF Astronomy \& Astrophysics Postdoctoral Fellow
\and
Instituto de Astrof\'{\i}sica de Canarias, Calle V{\'i}a L{\'a}ctea s/n, E-38205 La Laguna, Tenerife, Spain
\and
Department of Astronomy, The Ohio State University, 140 West 18th Avenue, Columbus, OH 43210, USA 
\and
National Optical Astronomy Observatories, Tucson, AZ 85719, USA
\and
Departamento de Astrofísica, Universidad de La Laguna (ULL), E-38206 La Laguna, Tenerife, Spain 
\and
Department of Astronomy, University of Virginia, Charlottesville, 3530 McCormick Road, VA 22904-4325, USA 
\and
Leibniz-Institut für Astrophysik Potsdam (AIP), An der Sternwarte 16, 14482 Potsdam, Germany
\and
Laborat\'{\o}rio Interinstitucional de e-Astronomia, - LIneA, Rua Gal. Jose\'e Cristino 77, Rio de Janeiro, RJ - 20921-400, Brazil
\and
Apache Point Observatory and New Mexico State University, P.O. Box 59, Sunspot, NM, 88349-0059, USA 
\and
Sternberg Astronomical Institute, Moscow State University, Moscow
\and
University of Texas at Austin, 32 Fowlkes Rd, McDonald Observatory, Tx 79734-3005 
 }


%

\abstract {The  inner Galactic Bulge  has,  until recently, been avoided in  chemical evolution studies due to extreme extinction and stellar crowding. Large, near-IR spectroscopic surveys, such as APOGEE, allow for the first time the measurement of metallicities  in the inner region of our Galaxy. We study  metallicities of 33 K/M giants situated in the Galactic Center region from observations obtained with the APOGEE survey.  We selected K/M giants with reliable stellar parameters from the APOGEE/ASPCAP pipeline. Distances, interstellar extinction values, and radial velocities  were checked to confirm that these stars are indeed situated in the inner Galactic Bulge. We find a metal-rich population centered at [M/H] = +0.4\,dex, in agreement  with earlier studies of other bulge regions, but also a peak at  low metallicity around  $\rm [M/H] = -1.0\,dex$, suggesting the presence of a metal-poor population which has not previously been  detected in the central region.  Our results indicate a dominant metal-rich population with a metal-poor component that  is enhanced in the $\alpha$-elements. This metal-poor population may be associated with  the classical bulge and a fast formation scenario.}


\maketitle

\titlerunning{ }
\authorrunning{ }

\section{Introduction}
 The Milky Way bulge is such a complex system that its formation and evolution are still poorly understood. Due to the high extinction and crowding, the study of the Galactic bulge remains challenging. Extinction of more than 30\,mag in $\rm A_{V}$  in the Galactic Center (GC) regions requires IR spectroscopy. 
While more and more detailed chemical abundances in the intermediate
and outer Bulge (such as Baade's Window) are now available thanks to the large spectroscopic surveys ARGOS (\citealt{freeman13}), Gaia-ESO (\citealt{rojas2014}), and APOGEE (\citealt{garciaperez2013}), chemical abundances of stars in the Inner Galactic Bulge (IGB) with projected distances of $\rm R_{G} \leq 200\,pc$ from the GC remain poorly studied. 

Most of the chemical abundance studies in the IGB  have been limited to luminous supergiants whose complex and extended stellar atmospheres make abundance analysis difficult.  \citet{Carr2000}, \citet{Ramirez2000},  \citet{cunha2007}, and \citet{Davies2009}  analysed high-resolution spectra of supergiant stars in the GC region and found  metallicities of near solar metallicity. A similar conclusion was obtained by \citet{Najarro2009} after analysing two luminous blue variables. \cite{cunha2007}  derived abundances nine giant and supergiant stars in the Central Cluster and found a metallicity of $\rm [Fe/H]= +0.14 \pm 0.06$, together with enhanced [O/Fe] and [Ca/Fe] abundances. The total range in Fe abundance among GC stars, 0.16 dex, is significantly narrower than the iron abundance distributions found in the literature for the older bulge population. Ryde \& Schultheis (2015, RS2015) studied
nine field stars  with a projected distance $\rm R \leq 50\,pc$. They found a metal-rich population with  $\rm [Fe/H]= +0.11 \pm 0.15 $ and low $\alpha$-value, and a lack of a metal-poor population, similar to \citet{cunha2007}.  Their mean metallicities are $\sim$ 0.3\,dex higher than fields in the inner bulge (\citealt{rich2007}, \citealt{rich2012}), indicating that the GC region contains  a distinct population. \citet{grieco2015} compared these data with a chemical evolution model and concluded that in order to reproduce
 the observed $\rm [\alpha/Fe]$ ratios, the GC region should have experienced a main strong burst of star formation and evolved very quickly with an IMF  which contains more massive stars.  
 All these studies comprise a  limited number of targets with precise chemical abundances. In this letter, we will discuss  the metallicity distribution  of 33 APOGEE K/M giants in the so-called ``GALCEN field'', increasing significantly the statistics of K/M giant stars with reliable metallicities.


\section{The Sample} \label{sec:sample}

\subsection{APOGEE}
The Apache Point Observatory Galactic Evolution Experiment (APOGEE; \citealt{majewski2012}, 2015 in prep.), one of four Sloan Digital Sky Survey-III (SDSS-III, \citealt{eisenstein2011}) experiments, is a large scale, near-IR,  high-resolution ($R \sim 22,500$) spectroscopic survey of Milky Way 
stellar populations. The survey uses a dedicated, 300-fiber, cryogenic spectrograph that is 
coupled to the wide-field, Sloan 2.5\,m telescope (\citealt{gunn2006}) at the Apache Point Observatory (APO).
APOGEE observes in the $H$-band, 
where extinction by dust is significantly lower than at optical wavelengths 
(e.g., $A(H) / A(V) \sim 0.16$). 

With its high resolution and high S/N ($\rm \sim 100$ per Nyquist-sampled pixel),
APOGEE determined both accurate radial velocities (better than $\rm 0.5\,km\,s^{-1}$) and
reliable  abundance measurements, including the most abundant metals in the universe (C, N, O), along with other $\alpha$, odd-$Z$, and iron-peak elements.
The latest SDSS-III Data Release (DR12; \citealt{ahn2014})  provides spectra of more than 140,000 stars to the scientific community, as well
as the derived stellar properties, including radial velocities, effective temperatures, surface gravities, metallicities, and individual
abundances. Additional information, such as photometry and target selection criteria, is also provided and described in \citet{zasowski2013}.

\subsection{The GALCEN field} 
The GALCEN APOGEE field is centered at (l, b)= (0.173$\rm ^{o}$, {\mbox{-0.07$\rm ^{o}$}) with a field of view of  $\sim$ 2.9  sq. degree (see Fig.~\ref{coord}), and it consists of
  339 stars. Due to the high interstellar extinction (see, e.g., \citealt{schultheis2014}),  a special target selection procedure has been applied for the GALCEN field (\citealt{zasowski2013}). Forty eight stars are known to be Long-Period Variables based on K-band lightcurves  (\citealt{matsunaga2009}). Thirty four stars are  Asymptotic Giant Branch (AGB) stars based on low-resolution HK-band spectra (\citealt{schultheis2003}). In addition,  spectroscopically known  supergiants such as IRS7, IRS19, and IRS22 were observed as well as supergiant candidates based on
photometric criteria. These AGB and supergiant  targets have among the coolest effective temperatures ($\rm T_{eff} <   3500 K$) in the APOGEE stellar sample.
 Figure~\ref{coord} shows the distribution of the APOGEE GALCEN sample in Galactic coordinates. 
 Only $\sim 10\%$  of the 339 GALCEN stars (filled circles) were used for our analysis (Sect.~3). 

\begin{figure}[!htbp]
  \centering
	\includegraphics[width=0.49\textwidth,angle=270]{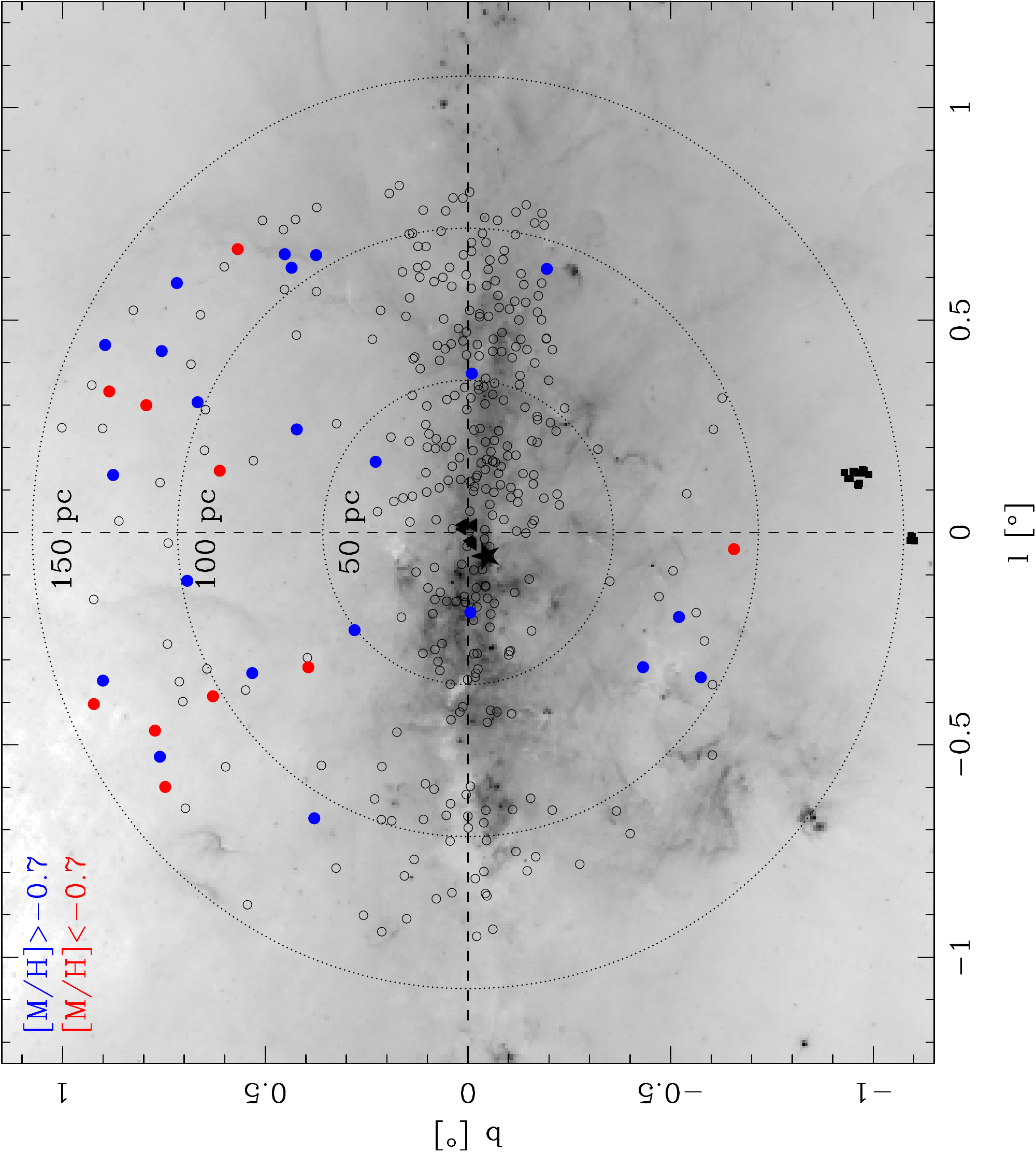}
	\caption{Galactic longitude vs. Galactic latitude distribution of the  GALCEN sample superimposed on the 8\,$\mu$m IRAC image of GLIMPSE-II (\citealt{churchwell2009}). The red filled symbols indicate our selected sample with $\rm [M/H] < -0.7$, the blue ones with $\rm [M/H] < -0.7$ and the open black symbols the  full GALCEN sample. Circles denote galactocentric radii of 50, 100, and 150\,pc, assuming $\rm R_{\odot} = 8\,kpc$. The black asterisk show the  GC field of Cunha et al. (2007), the black triangles those of  Ryde\&Schultheis (2015), black squares  those of  Rich et al. (2007).} 
	\label{coord}
\end{figure}

Figure~\ref{CMD} shows the dereddened  colour-magnitude diagram ($\rm (H-K_{0})$ vs. $K_{0}$)  of the selected GALCEN sources
(black points), together with the 2MASS  data in this field  that were dereddened using the  high-resolution VVV extinction map of \citet{schultheis2014}.
The GALCEN sample sources on the other hand, were dereddened using the RJCE method on a star-by-star basis  (see \citealt{zasowski2013} for more details).   Superimposed are the Padova isochrones with an age of 10\,Gyr and  $\rm [Fe/H]=-0.4$ and $\rm [Fe/H]=+0.3$.

\begin{figure}[!htbp]
  \centering
	\includegraphics[width=0.49\textwidth]{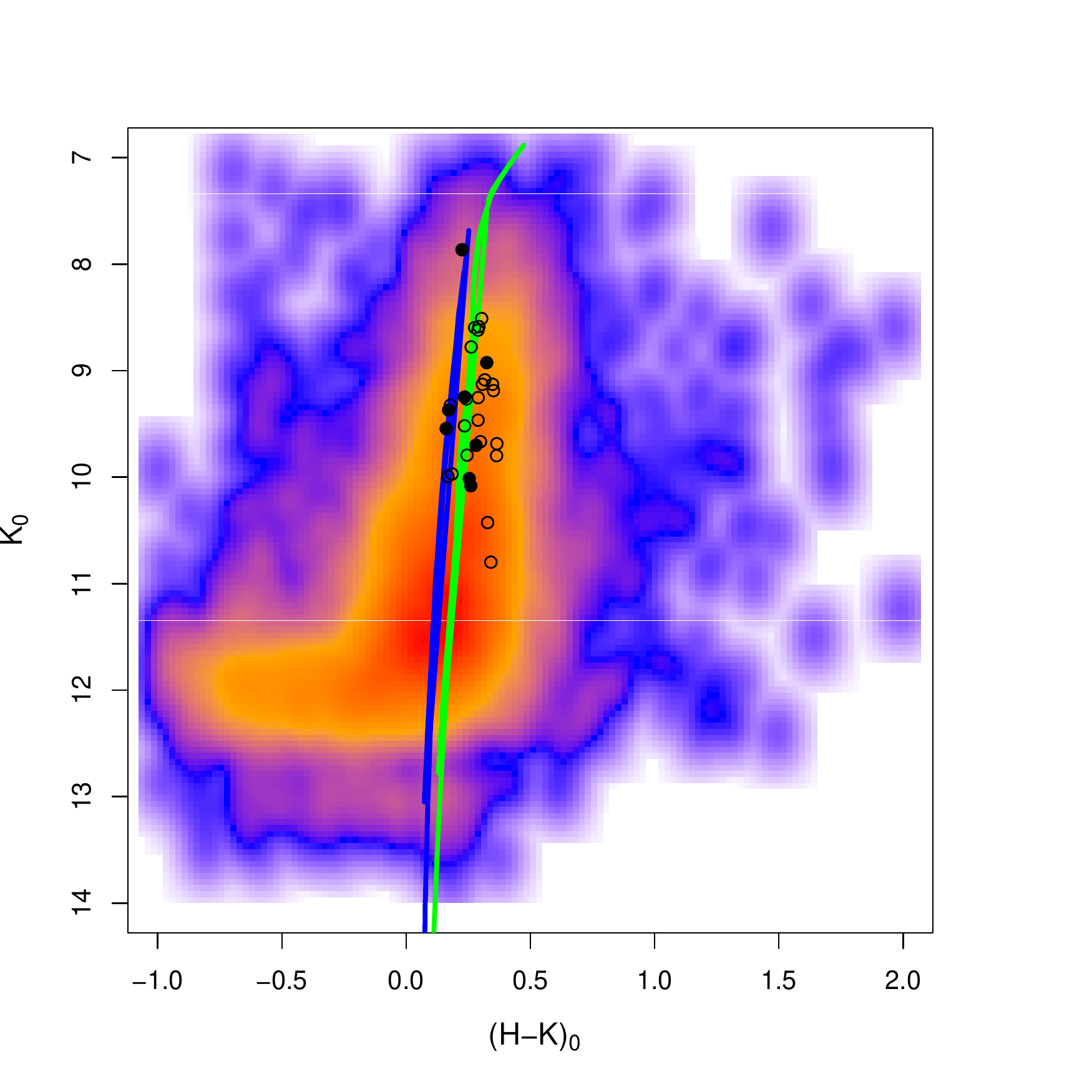}  
	\caption{Dereddened $\rm (H-K)_{0}$  vs. $\rm K_{0}$ 2MASS diagram using the extinction map of Schultheis et al. (2014). The blue line shows the Padova isochrones of 10\,Gyr with $\rm [Fe/H]=-0.4$   while the green line $\rm [Fe/H]=+0.3$.  Black circles show our sample of stars dereddened  with the RJCE method. Open circles show $\rm [Fe/H] > -0.7$ while the filled ones are the low metallicity stars with $\rm [Fe/H] < -0.7$.} \label{CMD} 
\end{figure}

\subsection{Stellar parameters} \label{parameters}
Stellar parameters and chemical abundances of up to 15 elements
are determined by the ASPCAP pipeline.
These values are based on a $\chi^{2}$-minimization between observed and synthetic model spectra (see \citealt{zamora2015}, \citealt{holtzman2015} for more details on the DR12 spectral libraries)
performed with the {\sc FERRE} code \citep[][and subsequent updates]{allende2006}.
Model spectra are interpolated on a regular grid computed with the ASS$\epsilon$T code \citep{koesterke2009} for
a custom line list specially compiled for the survey (\citealt{shetrone2015}),
and  a large grid of model atmospheres (\citealt{meszaros2012}).

The accuracies of the  ASPCAP $T_{\rm eff}$, $\log{g}$, and [M/H] values were evaluated by \citet{holtzman2015} 
using a sample of well-studied field and cluster stars, including a large number of stars with asteroseismic stellar parameters from NASA's {\it Kepler} mission (\citealt{borucki2010}).  In this work, we use DR12 calibrated stellar parameters (we refer to \citealt{holtzman2015} for details).


\section{Sample selection} \label{selection}

In order to ensure the most reliable stellar parameters and abundances, we selected stars in the GALCEN field from the DR12 dataset with following criteria: (i) ASPCAP $\chi^{2} < 20$; (ii)   PARAMFLAG=0 for $\rm T_{eff}$, log\,g, and [M/H],  meaning that
  we avoid stars which are too close to the edge of the  model grid; and    (iii)  $\rm log\,{g} < 3$ to avoid foreground dwarfs. We exclude known supergiants, AGBs or possible AGB/supergiant candidates (based on near-and mid-IR criteria) for our analysis but restrict ourselves  to K/M giants.  In total we are left  with  33 K/M giants. We checked by visual inspection the reliabilty
of the  ASPCAP  fits to the observed spectra for each star.  Due to the cool effective temperatures of our stars (see Fig.~\ref{stelparams}), we restricted ourselves to using  temperatures, surface gravities, metallicities, and
global $\rm \alpha/Fe$ abundances, rather than the individual elemental abundances.  We added an additional cut in  effective temperature  ($\rm T_{eff} > 3700\,K$) for the analysis of the $\alpha$-elements to avoid possible systematic abundance effects in cool stars.



Figure~\ref{stelparams} shows the distribution of the sample stars  in  $\rm T_{eff}$ vs. $\rm log{g}$ space, colored by metallicity. One sees clearly that our sample
covers a wide metallicity range and that the majority of our targets are M giants.  As we exclude cooler stars with $\rm T_{eff} < 3600\,K $, we could be missing  more metal-rich stars.  In order to be sure that our stars are really located in the IGB,
 spectrophotometric distances were calculated using the Bayesian approach of \citet{anders2014}.  Our  stars are indeed located 
at a heliocentric distance  of  8\,kpc with a typical uncertainty in  the order of 20\%.  No difference in the distance distribution
was found between metal-rich and metal-poor stars.  In addition, our derived extinction values are compatible within 20\% with those of the IGB. We also studied
the radial velocity distribution of our sample, which has  a mean value of $\rm -32\,km\,s^{-1}$ with a dispersion of
$\rm 122\,km\,s^{-1}$. These values are similar to the velocity dispersion of $\rm 114\,km\,s^{-1}$ obtained by \citet{zhu2008}  for  the central parsec of the Galaxy  and to the results obtained in the study of \citet{rich2007},  with a  velocity dispersion of $\rm 115\,\pm 15\,km\,s^{-1}$. This is in perfect agreement with the predicted dispersion  by the N-body bulge model of \citet{shen2010}.
Interestingly, there is one metal-poor $\rm [Fe/H]=-1.0$ high radial velocity  star  with  $\rm v_{rad} = 356\,km\,s^{-1}$. Proper motions would be necessary to derive the orbital period of this star.

\begin{figure}
  \centering
        \includegraphics[width=0.49\textwidth]{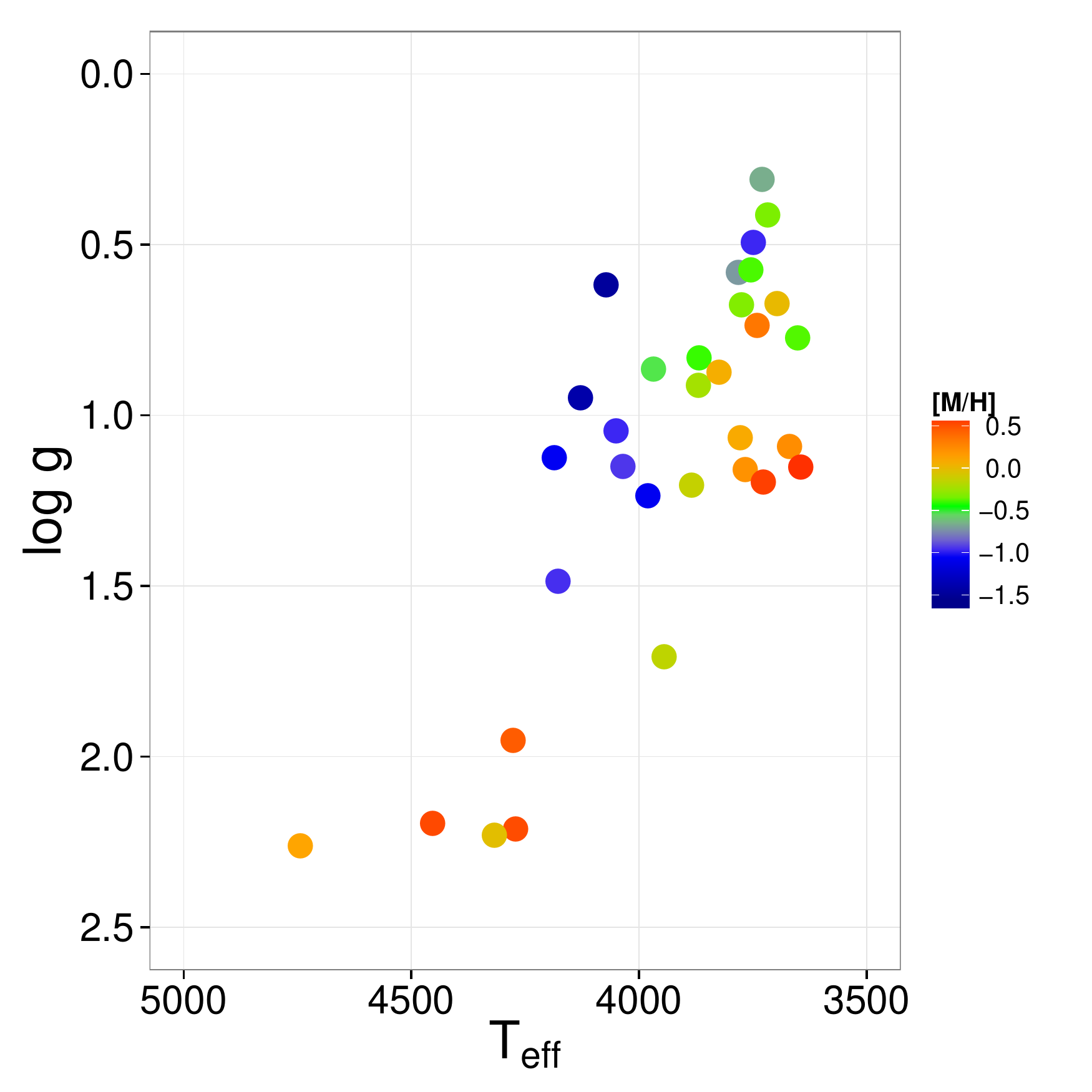}
	\caption{Teff vs. logg as function of [M/H] for our selected sample} 
	\label{stelparams}

\end{figure}

       

\section{Results and Discussion} \label{discussion}

Figure~\ref{histFeH} shows the histogram of the GALCEN sample metallicities. For comparison we show in green the sample
of \citet{rich2007} and in blue,  that of \citet{ryde2015}. We also show in red the metallicity distribution from the central
 cluster sample of \citet{cunha2007}, although this should be considered  a different stellar population. 
We clearly see a prominent peak in the metallicity distribution at $\sim$ 0.4\,dex, which is  about 0.2-0.3\,dex more metal-rich
 than found in \citet{ryde2015} and \citet{rich2007}.  The most striking feature in the metallicity distribution is the large dispersion, 
with a r.m.s. scatter of 0.55\,dex, which is much larger than seen in previous studies ($\rm \sim 0.15\,dex$),  and the presence of  a metal-poor peak at $\rm [Fe/H] \sim  -1.0$, which
has been not observed until so far in the IGB. This metal-poor population could be indicative  of a classical bulge and associated with a fast formation scenario (\citealt{babusiaux2010}).  A manual analysis of the $^{12}$C/$^{13}$C ratio in one of the metal-poor stars (2M17480557-2918376) reveals a value of $^{12}$C/$^{13}$C =5$\pm$2.  Such a low carbon isotope ratio for a red giant star is indicative of a low-mass first ascent giant (M$\sim$0.8 - 1.0 M$_{\odot}$)
as first quantified by \citet{gilroy1989} for open cluster giants and confirmed by \citet{mikolaitis2012}.  This value is close to that found in typical globular cluster red giants (e.g., \citealt{briley1994}) which typify an old,
low-mass population. 
Contrary to \citet{ryde2015}, \citet{rich2007}, and \citet{cunha2007}, the metallicity distribution
resembles a typical bulge population such as seen by \citet{bensby2013}, \citet{hill2011}, etc.  for bulge fields at higher latitudes ($\rm b >  4^{o}$). 
It should be noted that the metal-rich peak of our studied sample agrees  with the metal-rich peak of the galactic bulge sample of \citet{bensby2013} and the red clump bulge sample of \citet{hill2011}. However, the second peak in their metallicity distribution is at $\rm \sim -0.5$ to $\rm -0.6\,dex$  (compared to $\rm [M/H] \sim -1.0$ in our sample)  with a lack of  metal poor stars with $\rm [M/H] < -1$.
Our derived metallicity dispersion is  comparable
to  that of \citet{bensby2013} (0.53\dex) but larger than the  red clump distribution of \citet{hill2011} in Baade's Window.

\begin{figure}
  \centering
	\includegraphics[width=0.49\textwidth]{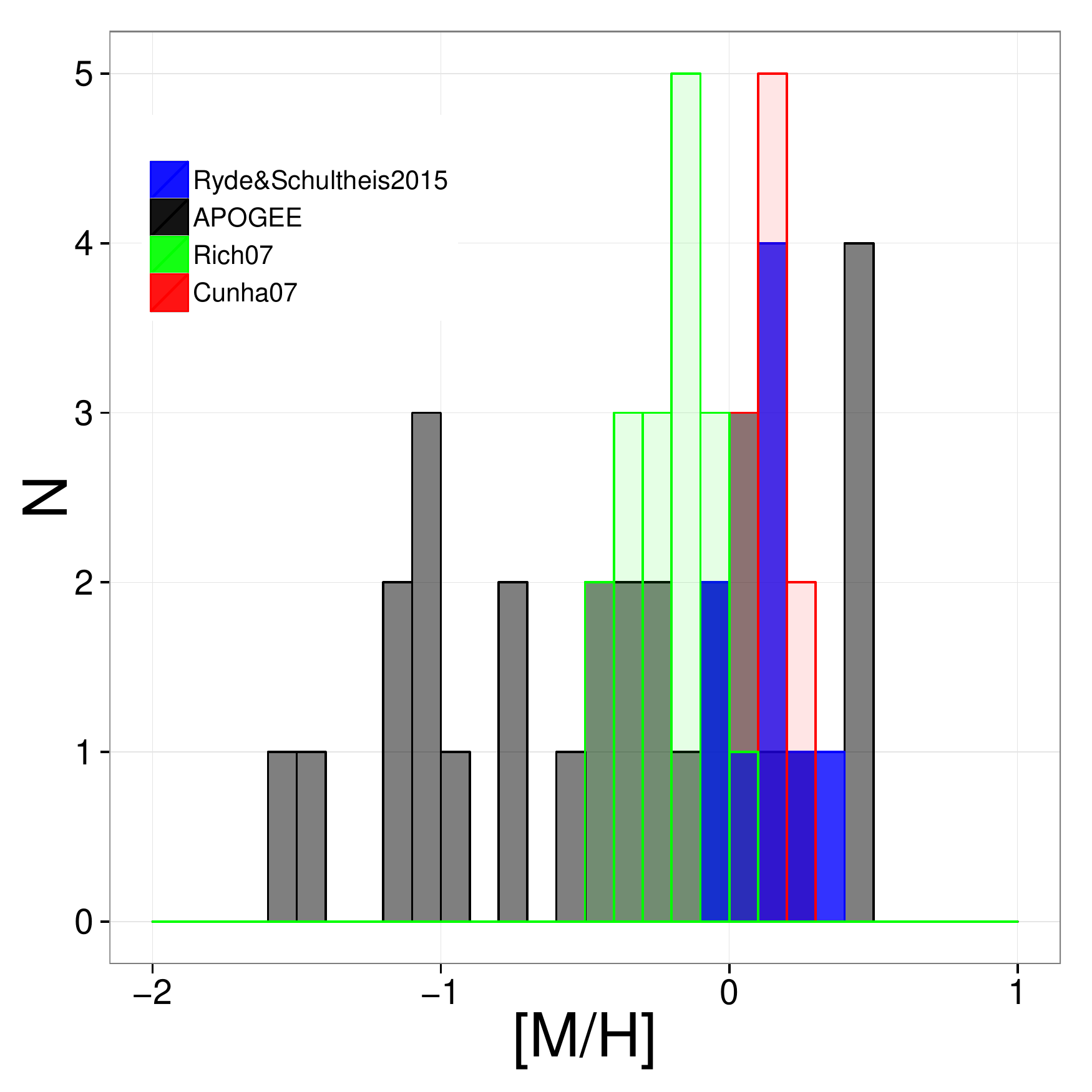}
     \caption{Histogram of the metallicities compared to \citet{rich2007} (green), \citet{cunha2007} (red) and \citet{ryde2015} (blue). The bin size is 0.1\,dex.}
       \label{histFeH}
        \end{figure}
What is the reason that this metal-poor population has not been revealed by other studies?
The \citet{ryde2015} and \citet{rich2007}   work contain  a much smaller sample, which could be one reason that they  have just missed these metal-poor stars. They deredden their
targets using global extinction maps such as \citet{schultheis1999}, \citet{schultheis2009}, or \citet{gonzalez2012}.   
On the contrary,  APOGEE calculates the reddening values on a star-by-star basis  using a combination of near
 and mid-IR photometry and applies a homogenous $\rm (J-K)_{0} > 0.5$ colour cut (see \citealt{zasowski2013} for further details). 

\begin{figure}
  \centering
	\includegraphics[width=0.49\textwidth]{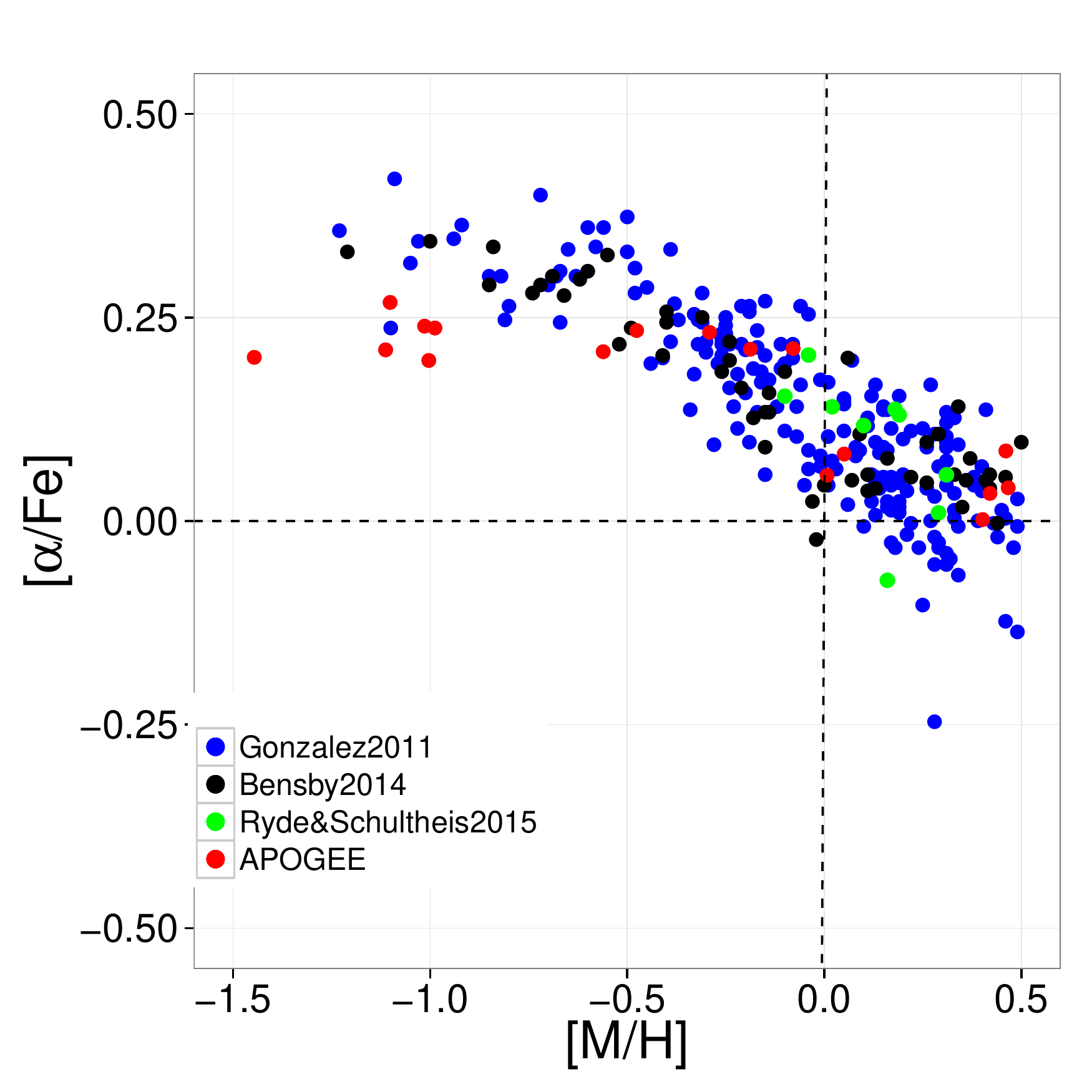}        
     \caption{$\alpha$ vs [M/H].  The blue filled circles indicate the galactic bulge data from \citet{gonzalez2011}, black from Bensby et al. (2013),  green from Ryde\&Schultheis (2015),  and in red  those of our GALCEN sources. We used only
stars with $\rm T_{eff} > 3700\,K$. }
        \label{abundances}
        \end{figure}

Figure~\ref{abundances} displays the $\alpha$-abundances of our  sample stars as a function of [M/H].  
As a comparison we also show the Galactic Bulge sample of  the dwarfs and subgiants stars of \citet{bensby2013}, the
M giants sample  in the Galactic Center of \citet{ryde2015}, and the red clump star sample in Baade's Window of \citet{gonzalez2011}. Our  $\alpha$-abundances agree  with those of \citet{ryde2015}, \citet{gonzalez2011} and \citet{bensby2013}  on the metal-rich side.   However, we notice that the $\alpha$-elements of \citet{bensby2013}  and \citet{gonzalez2011} are systematically higher than ours  for low-metallicity stars ($\rm [Fe/H] < -0.5$). 

None of the current models is able  to reproduce the observed metal-poor peak seen  in our  metallicity distribution. 
\citet{grieco2015} calculated for the first time a chemical evolution model for the
inner 200\,pc and compared it with the  dataset of \citet{ryde2015}. In order to produce their observed $\rm [\alpha/Fe]$ ratios and their metallicity distribution, they required a  model assuming a  main early strong burst of star formation
with a star formation efficiency of $\rm 25\,Gyr^{-1}$  and  a heavy IMF.  However, their MDF consists of a single metal-rich population.

Regarding the explanation for other metal-poor bulge peaks/populations, \citet{grieco2012}, for example,  used a chemical evolution model with two distinct populations that reproduces  the MDF of \citet{hill2011} in Baade's window. They concluded that  the  metal-poor population (which is different than the one discovered here)  was formed on a very short timescale associated with an intense burst of star formation and a high star formation efficiency, while the metal-rich population was formed on a longer time-scale and is related to the Galactic bar.  \citet{tsuji2012} modeled the MDF with a two-component disk model: a metal-poor component with a very short timescale of $\rm 1\,Gyr$ and identical to the solar-neighbourhood thick disc, and a metal-rich component with a longer timescale and a top-heavy IMF.

This shows that chemical evolution models can reproduce  observed MDFs but fail to predict  MDFs in different parts of our Galaxy. Thus, detailed confrontation with chemical evolution models are  necessary to understand the origin of the metal-poor peak.
 

\begin{acknowledgements}
Funding for SDSS-III has been provided by the Alfred P. Sloan Foundation, the Participating Institutions, the National Science Foundation, and the U.S. Department of Energy Office of Science. The SDSS-III web site is http://www.sdss3.org/. SDSS-III is managed by the Astrophysical Research Consortium for the Participating Institutions of the SDSS-III Collaboration including the University of Arizona, the Brazilian Participation Group, Brookhaven National Laboratory, Carnegie Mellon University, University of Florida, the French Participation Group, the German Participation Group, Harvard University, the Instituto de Astrofisica de Canarias, the Michigan State/Notre Dame/JINA Participation Group, Johns Hopkins University, Lawrence Berkeley National Laboratory, Max Planck Institute for Astrophysics, Max Planck Institute for Extraterrestrial Physics, New Mexico State University, New York University, Ohio State University, Pennsylvania State University, University of Portsmouth, Princeton University, the Spanish Participation Group, University of Tokyo, University of Utah, Vanderbilt University, University of Virginia, University of Washington, and Yale University.
\end{acknowledgements}


\bibliographystyle{aa}
\bibliography{apogee_galcen_accepted}

\end{document}